\documentclass{article}

\usepackage[final]{neurips_2023}

\usepackage{mathtools}

\usepackage[utf8]{inputenc} 
\usepackage[T1]{fontenc}    
\usepackage{hyperref}       
\usepackage{url}            
\usepackage{booktabs}       
\usepackage{amsfonts}       
\usepackage{nicefrac}       
\usepackage{microtype}      
\usepackage{xcolor}         
\usepackage[T1]{fontenc}    
\usepackage{hyperref}       
\usepackage{url} 
\usepackage{booktabs}       
\usepackage{amsfonts}       
\usepackage{nicefrac}       
\usepackage{microtype}      
\usepackage{lipsum}		
\usepackage{graphicx}
\usepackage{natbib}
\usepackage{doi}
\usepackage{amsmath}
\usepackage{graphicx}
\usepackage{caption}
\usepackage{subcaption}
\usepackage{multirow}
\usepackage{float}
\usepackage{caption}
\usepackage{subcaption}
\usepackage{wrapfig}
\usepackage{siunitx}
\usepackage{placeins}

\setcitestyle{numbers,square}

\setcitestyle{numbers,square}

\title{Pay Attention to Mean-Fields for Point Cloud Generation}

\author{
Benno Käch\\
Deutsches Elektronen-Synchrotron DESY,\\  Germany\\
\texttt{benno.kaech@desy.de} \\
\And
Isabell-A. Melzer-Pellmann \\
Deutsches Elektronen-Synchrotron DESY,\\  Germany\\
\texttt{isabell.melzer@desy.de} \\
\And
Dirk Krücker\\
Deutsches Elektronen-Synchrotron DESY,\\  Germany\\
\texttt{dirk.kruecker@desy.de} \\
}

\begin{document}
\maketitle

\begin{abstract}
   
Collider data generation with machine learning has become increasingly popular in particle physics due to the high computational cost of conventional Monte Carlo simulations, particularly for future high-luminosity colliders. We propose a generative model for point clouds that employs an attention-based aggregation  while preserving a linear computational complexity with respect to the number of points. The model is trained in an adversarial setup, ensuring input permutation equivariance and invariance for the generator and critic, respectively. To stabilize known unstable adversarial training, a feature matching loss is introduced. We evaluate the performance on two different datasets. The former is the top-quark \textsc{JetNet150} dataset, where the model outperforms the current state-of-the-art GAN-based model, despite having significantly fewer parameters. The latter is dataset 2 of the CaloChallenge, which comprises point clouds with up to $30\times$ more points compared to the first dataset. The model and its corresponding code are available at \url{https://github.com/kaechb/MDMA/tree/NeurIPS}.

\end{abstract}

\section{Introduction}
Machine Learning (ML) has a longstanding presence in High Energy Physics (HEP) with its applications becoming standard across all stages of HEP data analysis~\cite{mlinhep}. While supervised learning has predominantly been used, there is a growing interest in generative models due to the detailed Monte Carlo simulations (MC) typically utilised in HEP. They provide near-perfect representations of experimental measurements, but they demand vast computing resources~\cite{mccms}. As higher and higher luminosities are reached (e.g. at the High Luminosity LHC)~\cite{hllhc} this becomes a more and more pressing problem. A potential solution could be provided by generative models, as they are orders of magnitude faster during inference and have proven to perform well, e.g.\ in image generation~\cite{dallee,stylegan}. However, there is no universal solution for assessing whether the multidimensional features that the model is sampling contain the same correlations between the different features as the true model.
Attention-based aggregation methods have displayed competitive performance for a wide array of problems. The inclusion of a mechanism in the proposed model was motivated by this factor. However, the quadratic scaling of self-attention with the number of inputs does not allow direct application to large point clouds. Thus, our information aggregation relies on cross-attention to an artificial mean-field, resulting in a linear scaling of computational complexity. 
The performance is first evaluated on the \textsc{JetNet150}~\cite{jetnet} top-quark dataset, which comprises of PYTHIA simulated particles~\cite{pythia}, clustered to jets using the anti-$k_T$
  algorithm~\cite{antikt}. While PYTHIA jet generation is not the main computational bottleneck that needs resolution, there are known complex high-level analytical correlations (e.g. the relative invariant jet mass) which are difficult to capture. The metrics discussed in~\cite{jetneteval} also allow for convenient benchmarking with other models from the field.
  Subsequently, the application of the same model to calorimeter simulation is explored on the CaloChallenge Dataset 2~\cite{calochallenge2}. The calorimeter energy deposits are first converted to point clouds to make use of the sparsity which is present. Although this requires extra positional coordinates of the hits, this efficiently represents the data by excluding empty cells. 

\section{Related Work}
\label{sec:related}
\subsection{\textsc{JetNet150}}
\label{sec:datasetjetnet}
The \textsc{JetNet150} datasets, introduced by Kansal et al.~\cite{jetnet}, contain five datasets of about 180,000 samples each. For the sake of brevity, we will only discuss the dataset containing jets initiated by top quarks. However, it is anticipated that the latter contains the most intricate substructure among all the available jet datasets. For further clarification, information regarding particle features and high-level correlations, such as the invariant mass of the jet, can be found in~\cite{jetnet}. Key metrics, Wasserstein distances, Frechet Physics Distance (FPD) and Kernel Physics Distance (KPD), are employed as defined in~\cite{jetneteval}. Multiple models~\cite{NFbenno,pcdroid,transflow,diffusionjetnet1,diffusionjetnet2,epic,jetneteval} exhibit noteworthy performance on these datasets. We compare our model to EPiC-GAN by Buhmann et al.~\cite{epic} as it is the only model that scales up to the \textsc{JetNet150} dataset~\cite{JetNet150v2} using only particle multiplicity as a condition, and no other jet observables as input to the generator. 
\subsection{CaloChallenge}
The CaloChallenge dataset contains calorimeter showers simulated by GEANT4~\cite{mc5},
 paired together with the energy of the incoming particle. To transform these to point clouds, the voxel index is used as a coordinate. The spatial location is dequantised by adding noise distributed as $p(x)=\frac{(n_2-n_1)}{n_1+n_2}\cdot x$, where $n_2, n_1$ are the total number of hits in neighbouring bins in the training set and  $x\sim U([0,1])$. Other models have been proposed for these datasets~\cite{mikuni2023caloscore, amram2023calodiffusion}, but their adaptability to non-idealised detectors with non-regular geometry remains uncertain. Due to the unavailability of general metrics that allow meaningful comparison, the primary focus will be on presenting and briefly discussing our results. On this dataset there are on average $\sim 1600$ hits, with a maximum of up to $\sim 4000$ hits per shower, and a total of 6480 cells, scaling up the number of points to generate in a cloud by a factor of up to $\sim 30$.

\newpage

\section{Architecture}
\FloatBarrier
\label{sec:architecture}
\begin{wrapfigure}{r}{0.4\textwidth}
    \centering

    \includegraphics[width=0.5\textwidth]{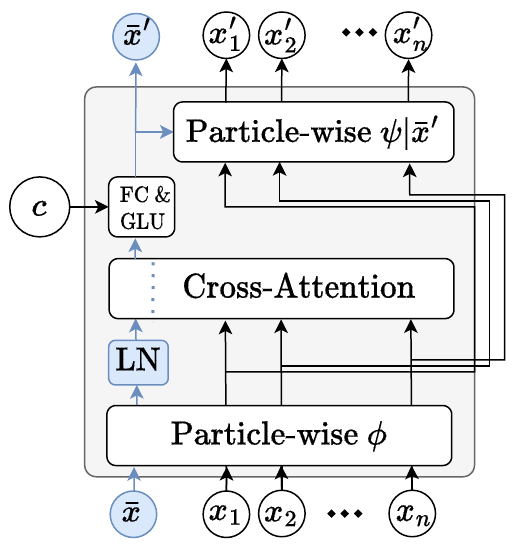}
    \caption{Minimal building block in the architecture comprises point-wise fully connected layers, $c$ are the variables that are used for conditioning using a gated linear unit, LN is a layer norm and FC stands for a fully-connected layer which is applied to the mean-field. The cross-attention layer is the sole means of interaction between different points.}
    \label{fig:critic}

\end{wrapfigure}
The model consists of a generator and a critic in a GAN setup. Both generator and critic accept a variable number of points as input and are built to be equivariant respective invariant under permutation of the input, as architectures that respect the underlying symmetries are a good treatment for the curse of dimensionality when optimising a function over a high-dimensional space~\cite{equivariance}. 
To aggregate information, a mean-field is introduced and initialized as the mean of all points. The interaction of all particles with each other is approximated by the interaction with this mean-field point only, allowing for scalability to large point clouds. The mean-field aggregates information via cross-attention to the points. The attention-based aggregation enables the critic and generator to dynamically select important points and disregard unimportant ones\footnote{Note that although attention is more commonly known from sequential NLP tasks, it is actually a permutation equivariant operation. It is therefore well suited to unordered sets such as point clouds. }
The minimal building block for the generator as well as the critic is depicted in Figure~\ref{fig:critic}. Multiple such blocks are stacked to build the main body of the architecture. Prior to the first block, a point-wise linear map is applied, mapping them to an $l$-dimensional latent space, in which further operations take place. The difference between the generator and the critic lies in the final layers of the architecture. 
\newline For the critic, a fully-connected neural network with one hidden layer is applied to the final mean-field to yield the score of the given input, whereas the generator maps the points down to the input dimension with a single point-wise layer.
Different methods of transferring information from the mean-field to the points were experimented with. The simplest technique is adding the mean-field to all other points. While this approach works to a certain extent, issues arise as this aligns all points with the mean-field making the calculation of the dot product attention problematic. A point-wise layer, that takes the points and the mean-field as input, comes at the cost of introducing more parameters, but its introduction is necessary for a competitive performance. 
Supplying the number of points in a shower explicitly to the model is crucial for the performance of the generator and critic. The model is implicitly conditioned on the number of points in a cloud $n$  by choosing the dimension of the input noise as $n\times f$ (where $f$ are the number of features per point). But since attention is a weighted sum, where the weights add up to unity, the mean-field is by construction agnostic to the number of elements during information exchange. To condition the showers with the incoming energy for the CaloChallenge dataset, a gated linear unit is utilised in the generator and critic.

\subsection{Training and Evaluation}
For the \textsc{JetNet150} dataset, we found that the LSGAN~\cite{LSGAN} training performs best, while the Wasserstein GAN together with gradient penalty~\cite{wgangp} performs best for the CaloChallenge dataset. Optimization of the network is done using Adam~\cite{adam} with momentum set to zero. The learning rate is varied with Cosine Annealing after a linear warmup to $10^{-4}$, which is commonly used with attention-based models~\cite{attention}. The critic for the CaloChallenge uses spectral normalisation~\cite{spectralnorm}\footnote{Note that this is done to enforce the Lipschitz constraint - however, it is shown in~\cite{l2attention} that the usual dot-product attention is not Lipschitz and they propose an L2-based attention, which is also applied in this work.}, while the one for \textsc{JetNet150} uses weight normalisation~\cite{weightnorm}.

\subsubsection{Deep Mean-field Matching}
To stabilise the generator training, we introduce a deep mean-field matching loss. In this approach the generator additionally optimizes the mean-squared error between the mean-field in the last layer of the critic for real and fake inputs. A similar loss is proposed by Goodfellow et al.~\cite{gantricks}, but they propose to minimize the squared difference of the mean. We found that this generally stabilises the training, however, the loss sometimes diverges later during training. Consequently, we limit the application of this loss for $50,000$ gradient steps. 

\section{Results \& Discussion}\label{sec:results}
In the following we summarize the most important aspects of the model, which we refer to as Matching Deep Mean-field Attentive (MDMA) GAN.
\subsection{\textsc{JetNet150}}
\begin{figure}
  \centering
  \includegraphics[width=\textwidth]{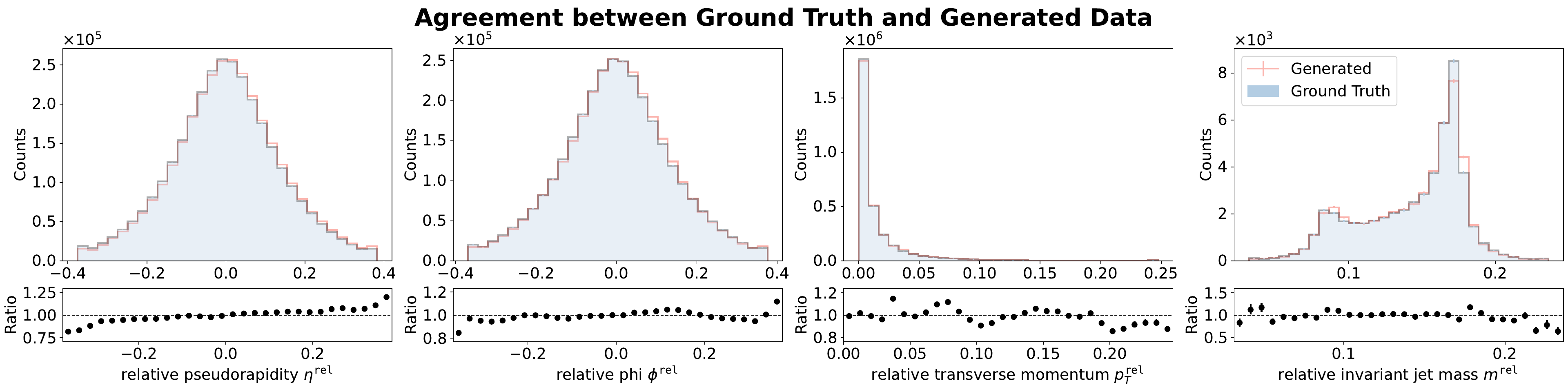}
  \caption{Comparison between the \textsc{JetNet150} dataset and generated samples. The first 3 plots shows the distribution of the marginal features $(\eta^{rel},\phi^{rel},p_T^{rel})$ of all particles. The plot on the right depicts the relative invariant jet mass which provides a difficult to model high-level correlation.
A ratio between the generated data and the ground truth is shown below every plot. }
  \label{fig:jetnet}
\end{figure}

During evaluation a kernel density estimate (KDE) is sampled to provide the condition of the number particles in a jet. The KDE is fit to the particle multiplicity distribution present in the training set.  In Table~\ref{tab:results}, the results are compared to the ones for EPiC-GAN. The in-sample metrics (IN) calculated between the training and testing sets are given for reference as well. To obtain the results from EPiC-GAN, the checkpoint provided in their repository was used. The MDMA-GAN model outperforms EPiC-GAN on every metric, albeit our generator/critic only has \num{111000}/\num{55100} parameters which is about a quarter compared to EPiC-GAN. Figure~\ref{fig:jetnet} depicts the marginal distributions and the relative invariant jet mass distribution for the true and generated data. 

\begin{table}[h]
\caption{Comparison between the model from Ref.~\cite{epic} (EPiC) and our model (MDMA) on the top-quark dataset. Bold font has been used to indicate scores where one model performs significantly better. The in-sample distance between the training and testing sample (IN) is also given for reference. Our model performs better on all metrics that were evaluated. The uncertainty on the metrics is given as the standard distance by bootstrapping the distances five times with a sample size of $50'000$.}
\label{tab:results}
\centering
\vspace{0.2cm}
\resizebox{\textwidth}{!}{%
\begin{tabular}{ll|lllll}
\hline
Jet Class & Model & $W_1^M (\times 10^{3})$ & $W_1^P (\times 10^{3})$ & $W_1^{EFP}(\times 10^{5})$ & $KPD (\times 10^{4})$ & $FPD(\times 10^{4})$ \\
\hline
\cline{1-7}\multirow{3}{*}{Top Quark}  &

MDMA & $\mathbf{0.4 \pm 0.1}$ & $\mathbf{2.5 \pm 0.1}$ & $\mathbf{2.0 \pm 0.3}$ & $\mathbf{-0.08 \pm 0.07}$ & $\mathbf{3.0 \pm 0.4}$ \\&
EPiC & $0.60 \pm 0.07$ & $3.79 \pm 0.09$ & $2.9 \pm 0.3$ & $2 \pm 1$ & $23 \pm 1$ \\&
IN & $0.3 \pm 0.1$ & $0.32 \pm 0.09$ & $1.0 \pm 0.3$ & $-0.1 \pm 0.1$ & $0.2 \pm 0.2$ \\\cline{1-7}

\end{tabular}%
}
\end{table}

\begin{figure}
  \centering
  \includegraphics[width=\textwidth]{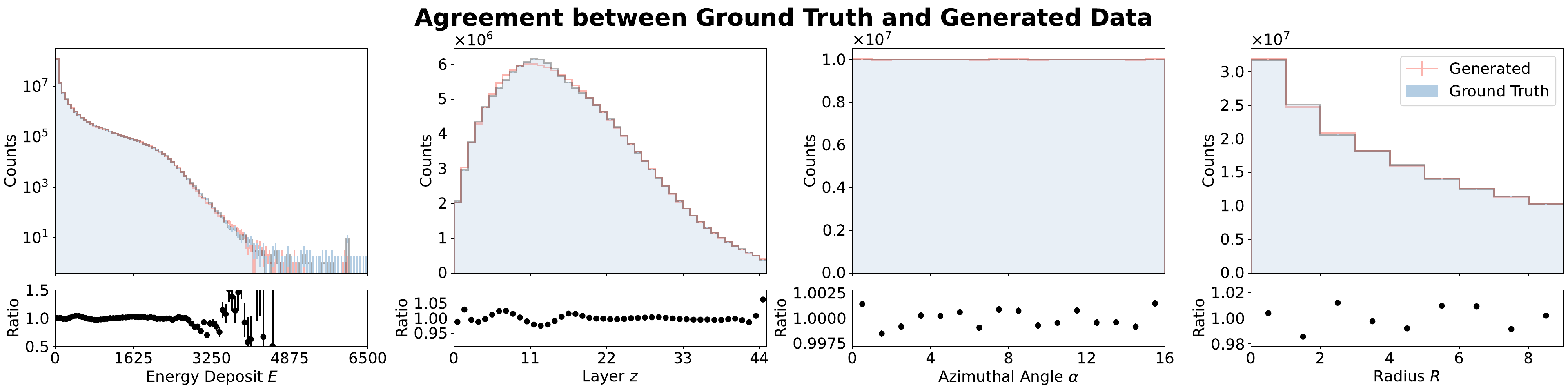}
  \caption{Distribution of the energy and position for MC simulated showers from the CaloChallenge dataset and for the ML generated showers.}
  \label{fig:calochallenge}
\end{figure}

\begin{figure}
    \centering
    \includegraphics[width=0.5 \textwidth ]{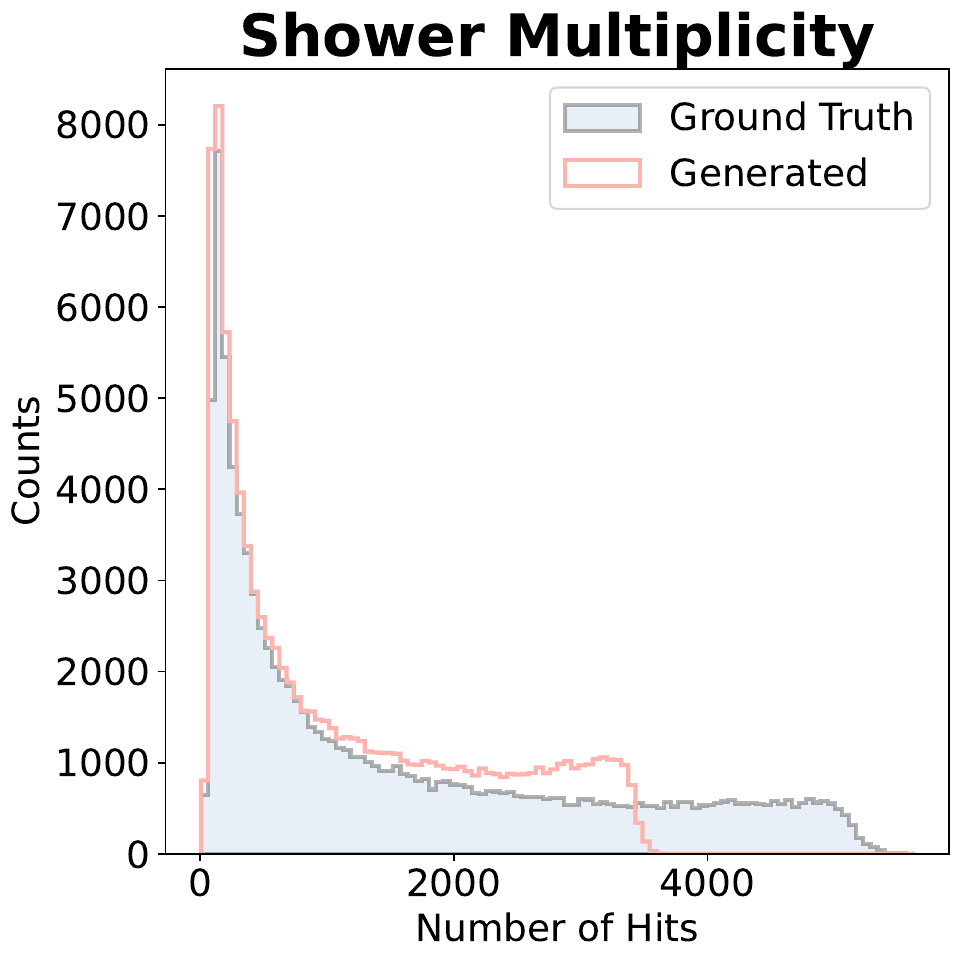}
    
    \caption{Limitation of point-cloud-based model: When transforming back to the voxel-based representation, numerous hits get mapped to the same cell leading to an incorrect distribution of the number of hits per shower.  }\label{fig:calochallenge_duplicates}

\end{figure}
\subsection{CaloChallenge}
Similarly, as for the \textsc{JetNet150} dataset, the marginal distributions of all hits are shown in
 Figure~\ref{fig:calochallenge}. Since ultimately the voxel-based representation of the data is needed, the shown results need to be considered with caution. When remapping the points to the voxel structure, multiple generated hits are often mapped back to the same cell as they are too close, which highlights a clear limitation of the model. This occurs predictably often for clouds containing a high number of points, as can be seen in Figure~\ref{fig:calochallenge_duplicates}.


\section{Conclusion}
We presented and evaluated a generative model for point clouds on two datasets. The model scales linearly with the number of points/tokens, thanks to the use of cross-attention to an artificial mean-field. To ensure stable adversarial training, the generator minimizes an L2 loss between mean-fields deep in the critic and the mean-fields from ground truth data. On the \textsc{JetNet150} top-quark dataset, the proposed model exhibits state-of-the-art performance using significantly fewer parameters than another state-of-the-art model with similar performance. \newline
Differences between in-sample distances and generated samples are nevertheless still observed. Furthermore, the model is deployed on the CaloChallenge dataset 2 where the point clouds yield up to $\sim 30$ more points to simulate. However, in this setting the model seems to be unable to model the spatial distribution of points correctly. 
It remains to be studied whether scaling up the model in terms of model size and amount of training data used resolves these remaining difficulties.
\newpage
\clearpage


\bibliographystyle{unsrtnat}


\bibliography{ref}  

\newpage




\end{document}